\documentclass[12pt,preprint]{aastex}

\shorttitle{Measurements of Solar Differential Rotation and Meridional Circulation}
\shortauthors{Imada \& Fujiyama}

\begin{document}

\title{Effect of Magnetic Field Strength on Solar Differential Rotation and Meridional Circulation}

\author{
S. \textsc{Imada},\altaffilmark{1} 
M.  \textsc{Fujiyama}\altaffilmark{1} 
}
  
\altaffiltext{1}{ Institute for Space-Earth Environmental Research (ISEE), Nagoya University, Furo-cho, Chikusa-ku, Nagoya 464-8601, Japan}

\begin{abstract}
We studied the solar surface flows (differential rotation and meridional circulation) using a magnetic element feature tracking technique by which the surface velocity is obtained using magnetic field data. 
We used the line-of-sight magnetograms obtained by the Helioseismic and Magnetic Imager aboard the Solar Dynamics Observatory from 01 May 2010 to 16 August 2017 (Carrington rotations 2096 to 2193) and tracked the magnetic element features every hour. 
Using our method, we estimated the differential rotation velocity profile. 
We found rotation velocities of $\sim$ 30 and -170 m s$^{-1}$ at latitudes of 0$^{\circ}$ and 60$^{\circ}$ in the Carrington rotation frame, respectively. 
Our results are consistent with previous results obtained by other methods, such as direct Doppler, time-distance helioseismology, or cross correlation analyses. 
We also estimated the meridional circulation velocity profile and found that it peaked at $\sim$12 m s$^{-1}$ at a latitude of 45$^{\circ}$, which is also consistent with previous results. 
The dependence of the surface flow velocity on the magnetic field strength was also studied. 
In our analysis, the magnetic elements having stronger and weaker magnetic fields largely represent the characteristics of the active region remnants and solar magnetic networks, respectively.
We found that magnetic elements having a strong (weak) magnetic field show faster (slower) rotation speed. 
On the other hand, magnetic elements having a strong (weak) magnetic field show slower (faster) meridional circulation velocity. These results might be related to the Sun's internal dynamics. 
\end{abstract}


\section{INTRODUCTION}

The 11-year variation in solar activity is an important source of decadal variation in the solar-terrestrial environment. 
The solar cycle variation is attributed to a dynamo occurring in the solar interior. 
Therefore, over the last few decades, considerable effort has been made to understand how the solar dynamo action occurs in the solar interior. 
To date, several theories have been proposed on the basis of both mean-field dynamo theory \cite[e.g.,][]{jou2008} and global three-dimensional magnetohydrodynamic simulations \cite[e.g.,][]{hot2016}. 
Although solar dynamo theory is still a matter of debate, the prediction of solar cycle activity has been intensively discussed recently in the context of space weather. 
The development of prediction schemes for the next solar cycle is a key to long-term space weather study, which is closely related to solar flare and coronal mass ejections \cite[e.g.,][]{tsu1992, ima2007, ima2011, ima2013}. 
The relationship between the polar magnetic field at solar minimum and the activity of the next solar cycle has received much attention in recent years. 
Many researchers currently believe that the polar magnetic field at solar minimum is one of the best predictors of the next solar cycle  \cite[e.g.,][]{sva2005}. 
To estimate the polar magnetic field, the surface flux transport (SFT) model has often been used, and several studies have succeeded in estimating the polar magnetic fields \cite[see][and references therein]{jia2014, iij2017}.
The SFT model requires several parameters such as the meridional circulation, the differential rotation, and the turbulent diffusion. 
These parameters have also been observationally investigated \cite[e.g.,][]{hat2011}.

Differential rotation, in which the solar surface rotates differentially depending on latitude, has long been discussed \cite[see][and references therein]{sch1985}. 
The rotation rate has been studied using Doppler measurements at the solar surface \cite[e.g.,][]{ulr1988}. 
The rotation rate has also been derived from the motion of magnetic features such as sunspots \cite[e.g.,][]{how1984} and magnetic elements \cite[e.g.,][]{kom1993}. Small-scale coronal features, \cite[EUV bright points,][]{bra2001}, \cite[X-ray bright points,][]{har2009}, also show differential rotation, although large-scale coronal features (for example, coronal holes) rotate almost rigidly. 
The differential rotation profile determined by coronal small-scale structures seems to correspond to the profile obtained from photospheric magnetic fields.

The meridional flow has also been discussed for a long time, and the poleward flow at low and intermediate latitudes is well established \cite[e.g.,][]{hat2014}. 
Because this flow is two to three orders of magnitude slower than the rotational flow, its basic entire structure is difficult to observe and is still controversial. 
It has been discussed whether or not the meridional flow varies with latitude, depth, and time. 
These variations in the flow are believed to be strongly related to the presence of sunspots; thus, the solar cycle is also believed to affect the meridional flow.
 
To date, various studies have investigated the basic structure of differential rotation and meridional circulation. 
However, few studies have focused on the influence of the magnetic field strength on the solar surface flow. 
We studied the solar surface flow, differential rotation, and meridional circulation using a magnetic element feature tracking technique in which the surface velocity is obtained using magnetic field data. 
To investigate the influence of the magnetic field strength on the solar surface flow, we also analyzed the relationship between the average magnetic field strength inside magnetic elements and the surface flow velocity.
  
\section{DATA AND OBSERVATIONS}

We use a series of line-of-sight magnetograms obtained from the Helioseismic and Magnetic Imager (HMI) aboard the Solar Dynamics Observatory \cite[SDO;][]{pes2012} from 01 May 2010 to 16 August 2017 (from CR2096 to CR2193, $\sim$100 Carrington rotations) at a cadence of 1 hr. 
The analyzed period corresponds roughly to the first half of solar cycle 24. 
We also calibrate the absolute value of the magnetograms using the method of \cite{liu2012}. 
The line-of-sight magnetic field is assumed to be largely radial, so we divide the magnetic field strength at each image pixel by the cosine of the heliographic angle from the disk center to minimize the apparent variations in field strength with longitude from the central meridian. 
Each full-disk magnetogram is mapped onto heliographic coordinates using the equidistant cylindrical projection \cite[e.g.,][]{kom1993,hat2011}. 
The resolution of the projected map is 0.1$^{\circ}$, and the range of the projection is $\pm$90$^{\circ}$ for the central meridional distance and latitude. 
To avoid a small signal-to-noise ratio for the magnetogram, we only use distances from the center of less than 75$^{\circ}$. 
Furthermore, we also use the Sun's rotation axis correction as described by \cite{hat2011}, in which they found that the accepted position of the Sun's rotation axis is in error by $\sim$0.08, as was noted previously by \cite{how1984} and \cite{bec2005}.

The magnetic element feature tracking technique has been discussed and is now well established \cite[e.g.,][]{iid2012,lam2017}. 
To detect the magnetic element features, we use a clumping method \cite[e.g.,][]{par2009} to identify each magnetic element having a magnetic strength exceeding a given threshold. The adopted threshold of 40 G was obtained by fitting the histogram of the magnetic strength. The magnetic element features are selected when the total magnetic flux inside the magnetic element ($\Phi$) is larger than 10$^{19}$ Mx. 
These threshold values were obtained from an evaluation of the noise level in the Michelson Doppler Imager (MDI) on the Solar and Heliospheric Observatory \cite[SOHO;][]{par2009}. 
It is well known that magnetic elements near sunspot behave differently. 
Thus, magnetic elements close to sunspots were masked out in a previous study \cite[][]{kom1993}. 
We define magnetic elements having a total magnetic flux larger than 10$^{21}$ Mx as sunspots and mask out magnetic elements less than 100 Mm from these defined sunspots. 
After the magnetic elements are detected, we track their motions. 
The travel distance of magnetic elements in a 1-hr interval is roughly 0.4$^{\circ}$--0.7$^{\circ}$, because the solar rotation speed is 10--15 deg day$^{-1}$. 
Therefore, we identify the same magnetic elements between two images to be within -0.1$^{\circ}$--1.0$^{\circ}$ in the longitudinal direction and -0.3$^{\circ}$--0.3$^{\circ}$ in the latitudinal direction. 
The detection of the merging and splitting of magnetic elements is generally difficult and has a high degree of uncertainty \cite[e.g.,][]{sch1997,iid2012}. 
To avoid this uncertainty, we track only the elements for which the total magnetic flux changes little. 
We define the flux change ratio (FCR) as $|$ log10($\Phi_2$/$\Phi_1$) $|$, where $\Phi_1$ and $\Phi_2$  are the magnetic flux of the magnetic elements in the previous and following maps, respectively. 
The magnetic elements are tracked when FCR $<$ 0.1. 
When there are several candidates, we select the element that has the lowest FCR value.

Figure 1 shows an example of our detection of magnetic elements. 
The full-disk magnetogram on 9 August 2010, mapped to heliographic coordinates using the equidistant cylindrical projection, is shown in Figure 1a. 
We detect and track only the magnetic elements inside the white dashed line in Figure 1a, which corresponds to 75$^{\circ}$. 
A total of 400 magnetic elements are detected by our method (Figure 1b). 
We identify 10 sunspots and mask out the magnetic elements close to them using the above criteria. 
Finally, we track 309 magnetic element features on this map (Figure 1c).
As seen in Figure 1, we identify the magnetic elements in both of the quiet sun and  the active region remnant.
Therefore, the magnetic elements having stronger and weaker magnetic fields largely represent the characteristics of the active region remnants and solar magnetic networks, respectively. 
We confirmed that our results, which are discussed below, are not sensitive to the detection criteria discussed above.

\section{RESULTS}

Figure 2 shows the average differential rotation profile derived from the entire data set from 01 May 2010 to 16 August 2017 ($\sim$100 CR). 
The velocities are taken relative to the Carrington frame of reference, which has a sidereal rotation rate of 14.184 deg day$^{-1}$. 
The average differential rotation profile is well fitted by the following equation:
\begin{equation}
f(\theta)=(a+b\sin^2(\theta)+c\sin^4(\theta))\cos(\theta),
\end{equation}
where
\begin{equation}
a=32.2 {\rm ~m~s^{-1}},
\end{equation}
\begin{equation}
b=-262.6 {\rm ~m~s^{-1}},
\end{equation}
\begin{equation}
c=-304.2 {\rm ~m~s^{-1}}.
\end{equation}
For comparison, we also added the fitted curve of \cite{hat2011}. 
We found rotation velocities of $\sim$30  and -170 m s$^{-1}$ at latitudes of 0$^{\circ}$ and 60$^{\circ}$, respectively. 
As shown in Figure 2, we can see a weak north-south asymmetry in the deviation of the measured profile from the symmetric profile (given by the red fitted line). 
The differential rotation was slightly stronger in the north than in the south. 
Flattening of the profile at the equator is also observed, which has been discussed in past studies \cite[e.g.,][]{sno1983}.

The meridional flow is generally more difficult to measure and has larger uncertainties than the rotational flow, because the meridional flow is two to three orders of magnitude slower than the rotational flow. 
The average meridional flow profile for the entire data set is shown in Figure 3. 
The profile at low latitude is well represented with just two antisymmetric terms illustrated in the following equation:
\begin{equation}
f(\theta)=(d\sin(\theta)+e\sin^3(\theta))\cos(\theta),
\end{equation}
where
\begin{equation}
d=30.1 {\rm ~m~s^{-1}},
\end{equation}
\begin{equation}
e=-26.4 {\rm ~m~s^{-1}} .
\end{equation}
However, the average meridional flow profile at a high latitude shows substantial differences from the fitted curve. 
The peak poleward meridional flow velocity is $\sim$12 m s$^{-1}$ at a latitude of 45$^{\circ}$. 
Our average meridional flow profile shows substantially different flows in the north and in the south. 
The flow velocity is faster in the north and peaks at a higher latitude than in the south. 
The flow in the south appears to nearly vanish at the extreme southern limit of our measurements, whereas the flow in the north is still poleward with a speed of about $\sim$5 m s$^{-1}$ at the northern limit.

We also study the influence of the magnetic field strength on the solar differential rotation and meridional circulation. 
Figure 4 shows the magnetic field strength dependence of the solar surface flow analyzed using the entire data set. 
The vertical axes in Figure 4a and b show the residuals of the differential rotation speed and meridional flow from the average at each latitude, respectively. 
The horizontal axes show the average magnetic field strength inside the magnetic element. 
Note that the field strengths that we are interested in are not the intrinsic field strengths of small-scale magnetic elements, which are likely to be kilogauss, because we average the field strengths inside the magnetic elements.
The data at each magnetic field strength are averaged, and the standard errors are shown (black lines: averaged with 10 data points). 
We can see that magnetic elements having a strong (weak) magnetic field show faster (slower) rotation speeds. 
On the other hand, magnetic elements having a strong (weak) magnetic field show slower (faster) meridional circulation velocities.
 
\section{Discussion and Summary}

We studied the differential rotation and meridional circulation using a magnetic element feature tracking technique and magnetic field data from 01 May 2010 to 16 August 2017 (approximately half of cycle 24). 
Using our method, we derived the differential rotation velocity profile and found rotation velocities of $\sim$30 and -170 m s$^{-1}$ at latitudes of 0$^{\circ}$ and 60$^{\circ}$ in the Carrington rotation frame, respectively. 
We also estimated the meridional circulation velocity profile and found that it peaked at $\sim$12 m s$^{-1}$ at 45$^{\circ}$. 
The dependence of the surface flow velocity on the magnetic field strength was also studied. 
We found that magnetic elements having a strong (weak) magnetic field showed faster (slower) rotation speeds. 
On the other hand, magnetic elements having a strong (weak) magnetic field showed slower (faster) meridional circulation velocities.

Let us discuss the validity of our results by comparison with past observations. 
We studied the differential rotation velocity in cycle 24. 
The angular rotation rate is nearly identical to that found by \cite{hat2011} for the time interval from 1996 to 2010 (mainly cycle 23) using different methods (a cross-correlation technique). 
The dashed line in Figure 2 shows the differential rotation result of \cite{hat2011}. 
The weak north-south asymmetry of the differential rotation and the flattening of the profile at the equator seem to be the same as those in our results. 
We also studied the meridional circulation velocity in cycle 24. 
The meridional circulation is also similar to that found by \cite{hat2011} for the time interval from 1996 to 2010 using different methods. 
Our meridional flow is slightly slower than the meridional flow discussed by \cite{kom1993} for the time interval from 1975 to 1991. 
\cite{hat2014} also discussed the meridional circulation in solar cycle 24 using HMI data from the time interval of April 2010 to July 2013. 
Their results also seem to be consistent with our results at not only low latitudes but also high latitudes. 
\cite{lam2017} also discussed the differential rotation and meridional circulation velocity using a magnetic feature tracking technique. 
Their results also seem to be consistent with ours, although their meridional circulation velocity at the peak latitude (45$^{\circ}$) is slightly faster than ours (16.7 m s$^{-1}$). 
This small discrepancy might arise from the difference in the observation period and the details of the tracking method. 
Therefore, we can conclude that our derived velocities are reasonable and consistent with past observations.

One of our interesting findings is the effect of the magnetic field strength on the solar differential rotation and meridional circulation. 
As mentioned in Section 2, we identify the magnetic elements in both of the quiet sun and  the active region remnant.
The magnetic elements having stronger and weaker magnetics field represent the characteristics of the active region remnants and solar magnetic networks, respectively.
Therefore, it is plausible that the magnetic field strength dependence on solar differential rotation and meridional circulation corresponds to the velocity difference between the active region remnant and the solar magnetic network.

\cite{dik2010} pointed out that the surface flow velocities obtained by magnetic elements feature tracking are affected by the surface turbulent magnetic diffusion.
They discussed that the diffusive transport of magnetic elements away from the active latitude produce an artificial outward flow from the active latitude.
\cite{hat2011} quantitatively estimated its effect from the observation and concluded that the diffusive transport effect must be limited.
They also discussed the impact of the inflow toward active latitude on the meridional flow and found that the inflow is one of the origins of the long-term variation of meridional flow profile.
We also checked the temporal variation of differential rotation speed and meridional flow and confirmed the presence of the inflow toward the active latitude and the torsional oscillation of Cycle 24 (not shown here).
These signals seem to be weaker compared to those of \cite{hat2011} (Cycle 23). 
It is plausible that the poleward flows of active region remnants decelerated by the inflow toward the active latitude.
The active region remnants is preferentially located at a higher latitude than the active latitude, where the inflow toward the active latitude is opposite direction to the meridional flow.
This might increase the magnetic field dependence of meridional circulation.
The impact of the inflow on the differential rotation speed should be limited, because the inflow accelerates and decelerates the differential rotation to the same degree. 

The dependence of magnetic field strength on solar differential rotation and meridional circulation might be related to the Sun's internal dynamics. 
The internal rotation speed and meridional flow of the sun have been studied by helioseismology. 
\cite{tho2003} studied the internal rotation speed of the sun and discussed the presence of a near-surface shear layer located in approximately the outer 5\% of the solar radius. 
According to their study, the internal rotation speed increases linearly with depth. 
If we assume that magnetic elements having a stronger magnetic field represent deeper regions in the solar interior, the trend of our results is consistent with that of the helioseismology observation. 
The same can be said of the meridional circulation. \cite{che2017} studied the internal meridional flow profiles. 
They found that the flow velocity decreases linearly by 30\% with depth around 0.975 R$_{sun}$ and is almost constant from 0.975 to 0.95R$_{sun}$. 
The trend of the magnetic field dependence on the meridional circulation in our analysis is also consistent with the helioseismic observation. 
If we compare our results with those of the helioseismology study, we can speculate that magnetic elements having fields of 250 G, 100 G, and 80 G might correspond to the flows at 0.95, 0.975, and 1.0 R$_{sun}$, respectively. 
We believe that our findings are useful for understanding the magnetic field transport at the solar surface and the structure of the near-surface shear layer.

\acknowledgments  We would like to thank the referee for the useful comments which helped improving the manuscript. The authors thank H. Iijima, H. Hotta, and Y. Iida for fruitful discussions. This work was partially supported by the Grant-in-Aid for 17K14401 and 15H05816. This work was also partially supported by ISEE CICR International Workshop program, and the authors thank all members of the workshop. The Solar Dynamics Observatory is part of NASA's Living with a Star program.

\begin{figure}
\epsscale{1.}
\plotone{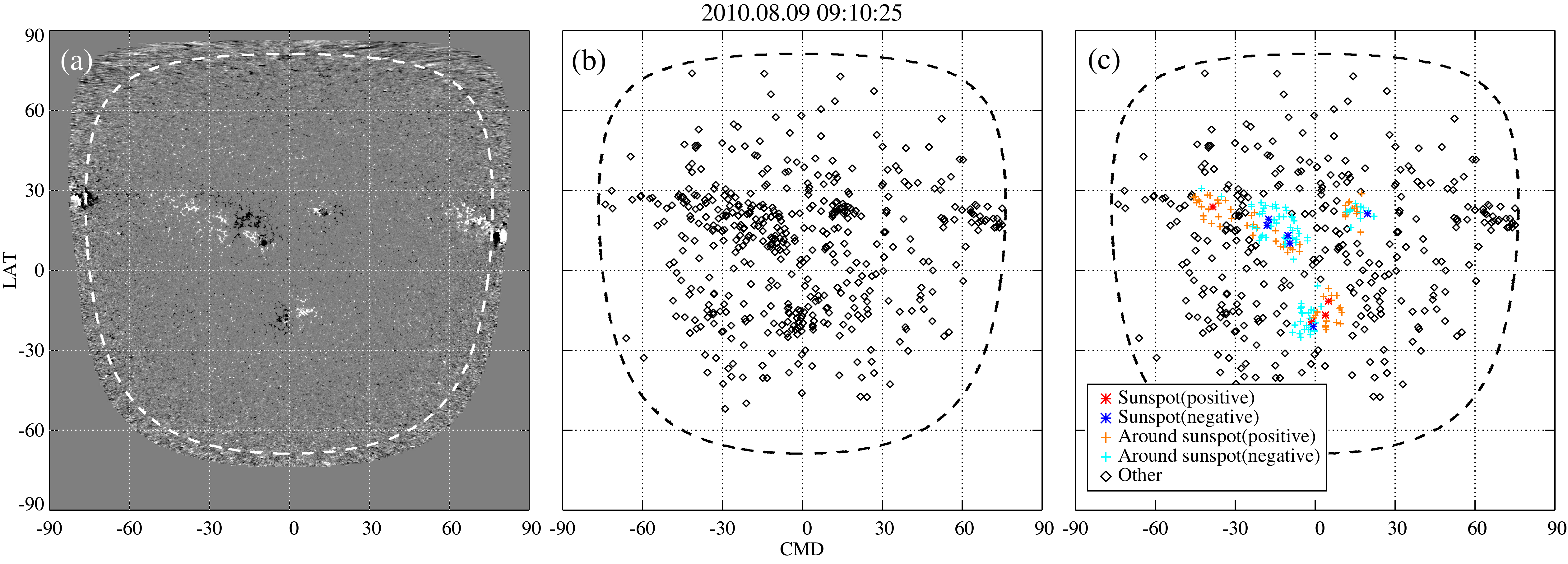}
\caption{
(a) Full-disk magnetogram mapped onto heliographic coordinates using equidistant cylindrical projection. 
(b) Magnetic element features detected by a clumping method. 
(c) Removal of magnetic elements that are close to sunspots. Red and blue asterisks represent positive and negative sunspots, respectively. 
Orange and sky-blue crosses are masked-out magnetic elements.
}
\end{figure}

\begin{figure}
\epsscale{0.7}
\plotone{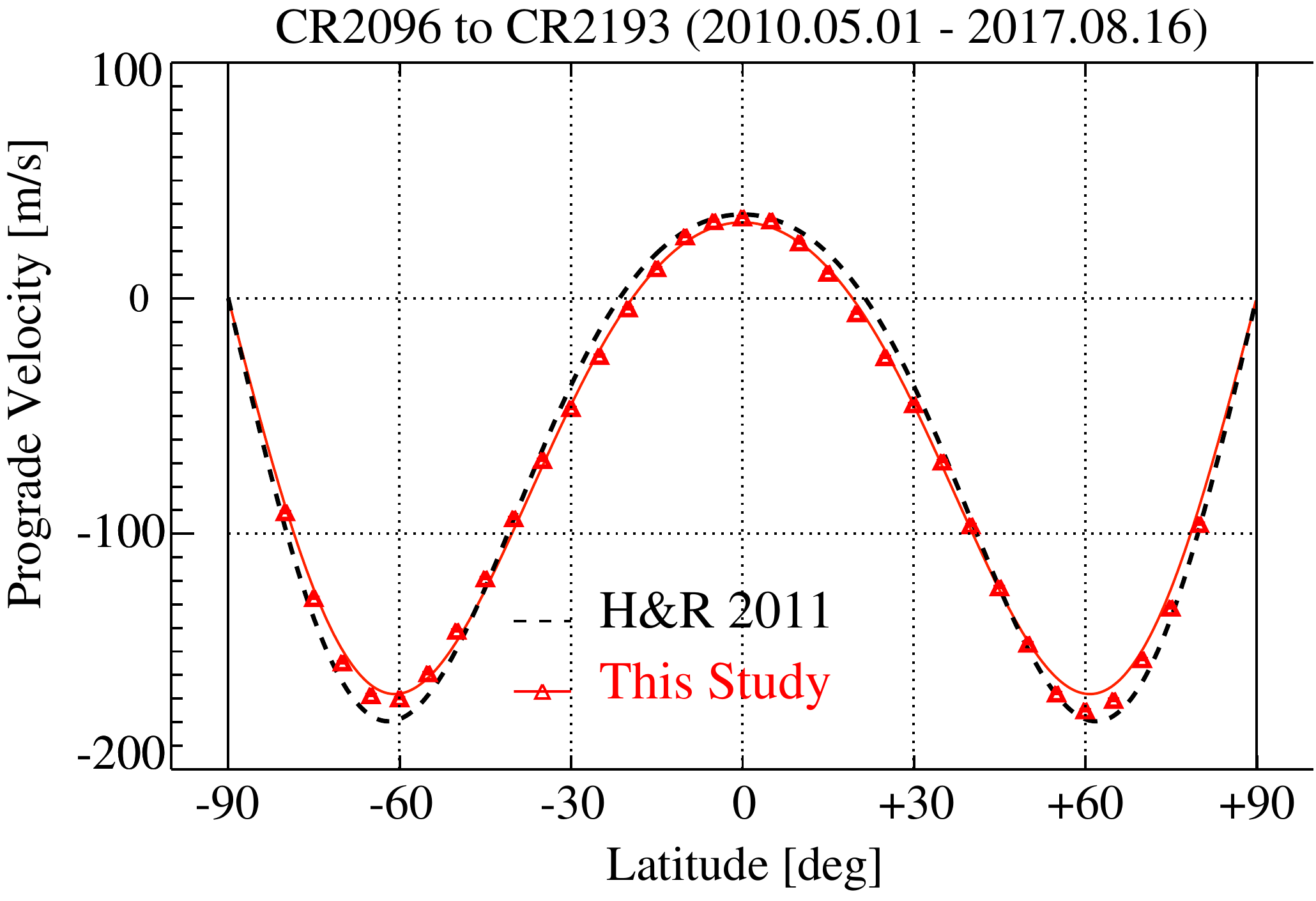}
\caption{
Average profile of differential rotation derived from the entire data set from 01 May 2010 to 16 August 2017 ($\sim$100 CR, half of cycle 24). 
Solid line: the fitted curve of this study, dashed line: the fitted curve of Hathaway \& Rightmire 2011. 
The prograde velocity values are taken relative to the Carrington rotation frame.
}
\end{figure}

\begin{figure}
\epsscale{0.7}
\plotone{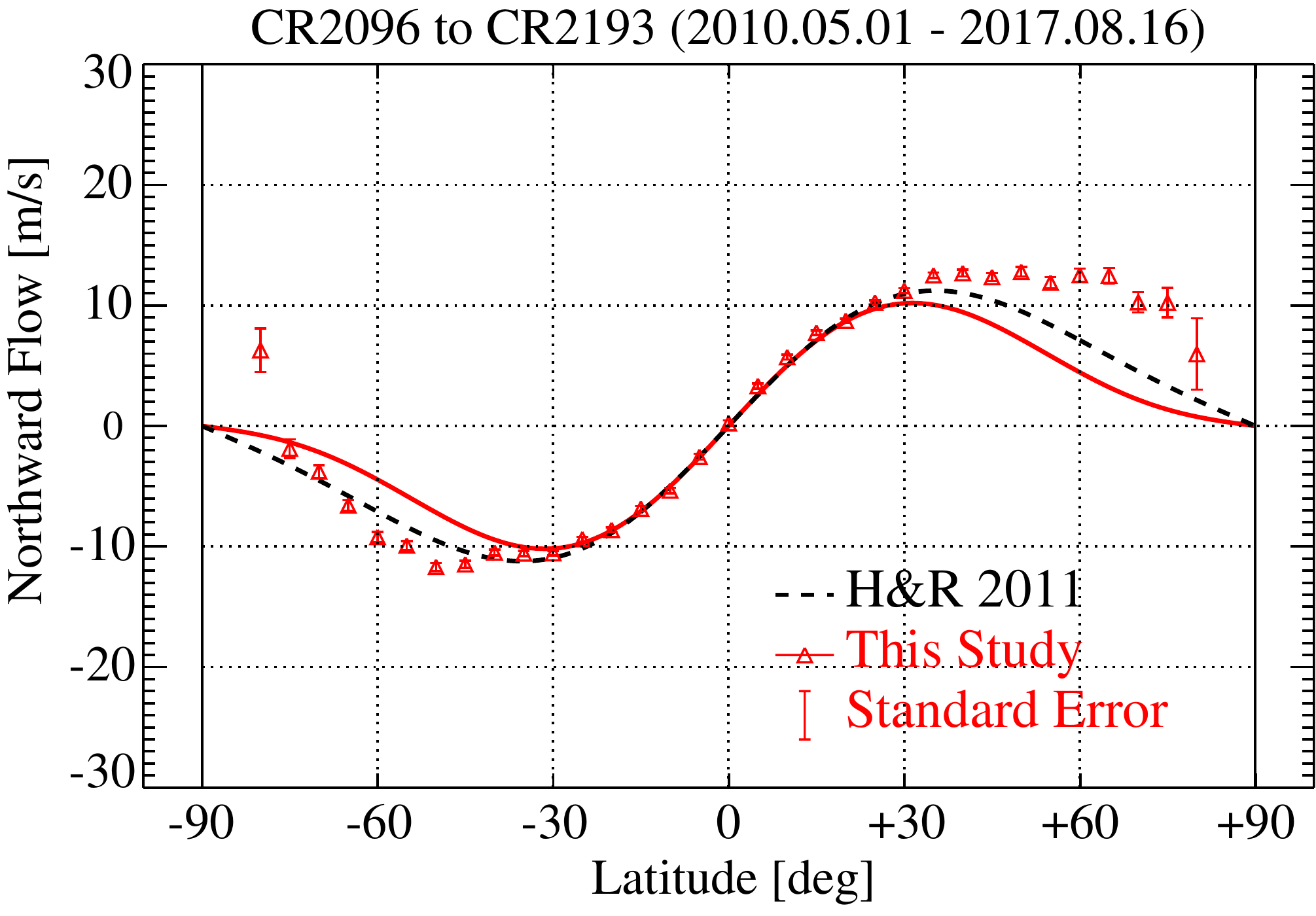}
\caption{
Average profile of meridional circulation derived from the entire data set from 01 May 2010 to 16 August 2017 ($\sim$100 CR, half of cycle 24). 
Solid line: the fitted curve of this study, dashed line: the fitted curve of Hathaway \& Rightmire 2011.
}
\end{figure}

\begin{figure}
\epsscale{1.0}
\plotone{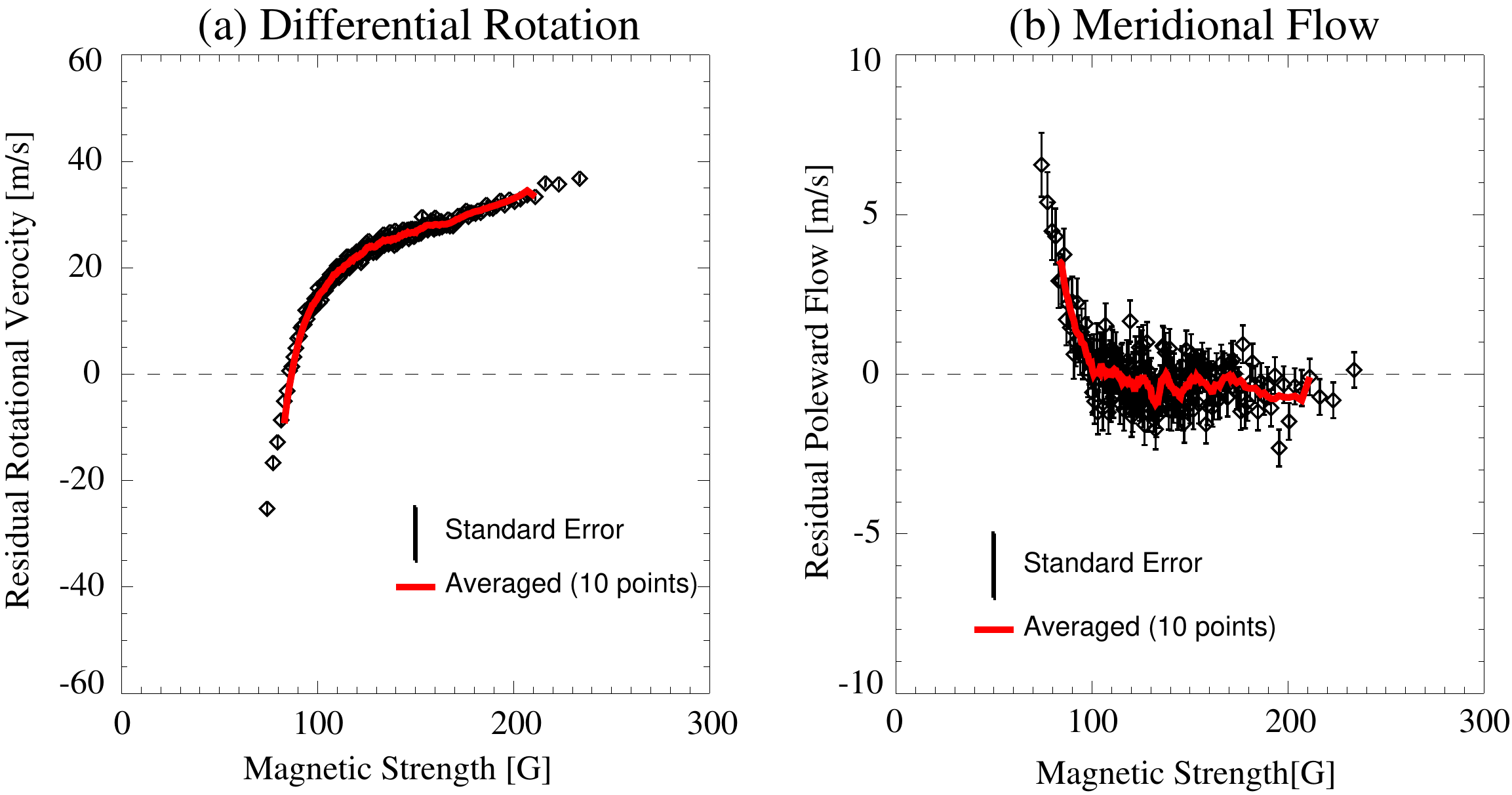}
\caption{ 
Effect of magnetic field strength on (a) solar differential rotation and (b) meridional circulation.
The vertical axes in Figure 4a and b show the residuals of the differential rotation speed and meridional flow from the average at each latitude, respectively. 
The horizontal axes show the average magnetic field strength inside the magnetic element.
}
\end{figure}

\end{document}